\documentclass[aps,12pt,eqsecnum]{revtex4}
\usepackage{amsmath}
\usepackage{graphicx}
\newcommand{\beq}{\begin{equation}}
\newcommand{\eeq}{\end{equation}}
\newcommand{\bes}{\begin{subequations}}
\newcommand{\ees}{\end{subequations}}
\newcommand{\bea}{\begin{eqnarray}}
\newcommand{\eea}{\end{eqnarray}}

\newcommand{\eps}{\mbox{\boldmath${\hat \epsilon}$}}

\begin{document}
\title{Quantum derivation of the use of classical electromagnetic potentials 
in relativistic Coulomb excitation}
\author{B.F. Bayman\\
{\it School of Physics and Astronomy, University of Minnesota,}\\
{\it 116 Church Street S.E., Minneapolis, MN 55455, U.S.A.}\\
and\\
F. Zardi\\
{\it Istituto Nazionale di Fisica Nucleare and Dipartimento di Fisica, \\}
{\it Via Marzolo,8 I-35131 Padova,Italy.}}
\begin{abstract}
We prove that a relativistic Coulomb excitation calculation in which the 
classical electromagnetic field of the projectile is used to induce 
transitions between target states gives the same target transition amplitudes, 
to all orders of perturbation theory, as would a calculation in which the 
interaction between projectile and target is mediated by a quantized 
electromagnetic field.
\end{abstract}
\maketitle
%\receipt{\today}
\date{\today}
\numberwithin{equation}{section}
\section{Introduction}
What has become the standard approach to relativisitic Coulomb excitation 
(RCE)
was proposed by A. Winther and K. Alder (WA) in 1979 \cite{WA}. 
The projectile nucleus is
assumed to travel along a straight-line orbit parallel to the ${\bf{\hat z}}$ 
axis, with impact parameter ${\bf b}$, at constant speed $v$.
The magnitude of the impact parameter is large enough so that nuclear
interactions between the target and projectile are negligible.  Because
of the assumed large projectile momentum, the electromagnetic impulse
the projectile receives due to its interaction with the target has
little effect on its trajectory, so the projectile maintains its
constant speed and impact parameter throughout the collision. As the
projectile passes, the target nucleus feels the time-dependent
projectile electromagnetic fields, which induce transitions between the
quantum states of the target. Starting from the assumption that the
target is in its ground state at $t=-\infty$, WA used
first-order perturbation theory to calculate the occupation probabilites
of excited target states at $t=+\infty$, and used these probabilities to
obtain Coulomb excitation cross-sections \footnote{Galetti, Kodama and Nemes 
\cite{GKN}, and Bertulani \textit{et al} \cite{BSMD}have pointed out diff
iculties associated with this simple picture.}.

An important ingredient in the calculation of target transition
probabilites is the interaction potential felt by the target as the
projectile moves past it. WA took this to be the
classical electromagnetic field of the moving projectile. A
justification of this assumption can be found in the work of Alder,
Bohr, Huus, Mottelson and Winther (ABHMW) \cite{ABHMW}. These authors used 
the
lowest-order of perturbation theory to calculate the target transition
amplitude as a result of photon exchange with the projectile, in a
situation in which the projectile motion was described by quantum
mechanics. They found that this photon-induced transition amplitude was
the same as the target transition amplitude induced by the classical
electromagnetic field of the projectile, again calculated in the lowest
order of perturbation theory.

It is well known \cite{HEI} that the quantum-mechanical treatment of the interaction of two charged particles yields the same result whether the interaction between them is
\begin{itemize}
\item The classical electromagnetic field, or
\item The quantized electromagnetic field, calculated up to terms of order 
$e^2$.
\end{itemize}
This is essentially the same result as reported by ABHMW\cite{ABHMW}.

In the 25 years since the WA paper, attempts have been made
to improve their calculation of transition amplitudes by going beyond
first-order perturbation theory \cite{BerBau,EML,ChFr,BCH,LCC,ABE,BZ,BZ2,BerPon99}. In particular, the time-dependent Schr\"odinger equation that governs the occupation amplitudes of the target states has been integrated numerically from $t=-\infty$ to
$t=\infty$ as a set of coupled-channel equations, within a finite set of
target basis states. In these calculations, the classical electromagnetic 
field has been
used as the interaction potential. But this clearly goes beyond the
justification provided by the work of ABHMW, which established the connection between the classical and quantized field only up to first-order perturbation theory.

In this paper we consider a somewhat restricted problem. We will assume 
that transition
charge and current densities that characterize the projectile are specified
 function of position and time, $\rho_{{\rm P}}({\bf r},t), {\bf J}_{{\rm P
}}({\bf r},t)$. In fact this is what is usually done in RCE calculations, with 
$\rho_{{\rm P}}({\bf r},t)$ and ${\bf J}_{{\rm P}}({\bf r},t)$ taken to
 be the charge and current densities appropriate to a spherically-symmetric
 charge moving with constant speed $v$ along a straight-line trajectory with 
impact parameter ${\bf b}$. We will not restrict the trajectory or the shape 
of the projectile; we only require that $\rho_{{\rm P}}({\bf r},t), {\bf J}_
{{\rm P}}({\bf r},t)$ be specified functions\footnote{Since we will neglect 
bremsstrahlung, our analysis only applies to situations in which the 
projectile does not suffer strong acceleration, which would certainly be the 
case in the constant-projectile-velocity scenario usually employed in RCE
.}. This means that we are neglecting the effects of projectile-state changes 
on the coupling of the projectile to the electromagnetic field. We will 
see that this commonly made assumption allows us to make a much stronger 
statement about the connection between the classical and quantized-field 
treatments of the excitation of the target. In fact, we will prove that they 
yield the same results to all orders of perturbation theory, not just to first 
order.

In Section II we will develop the basic expression for the transition amplitude in terms of the time-dependent interaction potential. In Section III 
we will evaluate this expression in the particular case in which the 
interaction is provided by the quantized electromagnetic field, calculated in 
the 
Coulomb gauge, and in Section IV we will show that this result is precisely
 the same as if we had used a classical electromagnetic field for the 
interaction. In Section V we will show that this agreement also holds 
if we had 
used the Lorentz gauge for both the classical and quantized field 
calculations. A more accurate treatment of the coupling between the projectile 
and the electromagnetic field is discussed in Section VI.
\section{The transition amplitude}

Consider the time-dependent Schr\"odinger equation
\beq
i\hbar\frac{\partial \psi}{\partial t}~=~[~h_0~+~v(t)~]~\psi
\label{2.1}
\eeq
where $h_0$ does not depend explicitly on $t$. This equation is expressed 
in the interaction
representation by expanding $\psi$ in terms of the normalized
eigenstates of $h_0$:
\beq
\psi(t)=\sum_\gamma
e^{-i\frac{E_\gamma}{\hbar}t}a_\gamma(t)~\Phi_\gamma,
\label{2.2}
\eeq
with
\bes
\bea
h_0\Phi_\gamma&=&E_\gamma\Phi_\gamma\\
\label{2.3a}
<~\Phi_\gamma~|~\Phi_\beta~>&=&\delta_{\gamma,\beta}
\label{2.3b}
\eea
\ees
If $\Phi_\alpha$ is the initial state, which existed when $t\rightarrow
 -\infty$, then
\beq
a_\gamma(-\infty)=\delta_{\gamma,\alpha}.
\label{2.4}
\eeq
The transition probability to state $\Phi_\gamma$ is given by
$|a_\gamma(+\infty)|^2$.

Because of equation (\ref{2.1}), the $a_\alpha(t)$ obey the set of
coupled
differential equations:
\bes
\beq
i \hbar \dot a_\gamma(t)=\sum_\beta~ e^{i
\Omega_{\gamma\beta}t}~ [v (t)]_{\gamma \beta} ~a_\beta(t),
\label{2.5a}
\eeq
with
\beq
\left[v(t)\right]_{\gamma \beta} ~\equiv~<\Phi_\gamma|v(t)|\Phi_\beta>.
\label{2.5b}
\eeq
\ees
$\Omega_{\gamma \beta}$ is defined by
\beq
\Omega_{\gamma \beta}\equiv \frac{E_\gamma-E_\beta}{\hbar}
\eeq
\label{2.6}

The integral equation equivalent to equation (\ref{2.5a}), incorporating
the initial condition (\ref{2.4}), is
$$
a_\gamma(t)=\delta_{\gamma,\alpha}+\int_{-\infty}^t~\frac{d t'}{i
\hbar}\sum_\beta~e^{i \Omega_{\gamma \beta}t'}v_{\gamma
\beta}(t')a_\beta(t').
$$
This can be iterated to develop a perturbation series in powers of
$v(t)$:
\beq
a_\gamma(t)=\delta_{\gamma,\alpha}+\int_{-\infty}^t\frac{d t'}{i
\hbar}e^{i \Omega_{\gamma \alpha}t'}v_{\gamma
\alpha}(t')+\sum_\beta\int_{-\infty}^t\frac{d t'}{i \hbar}e^{i
\Omega_{\gamma \beta}t'}v_{\gamma
\beta}(t')\int_{-\infty}^{t'}\frac{dt''}{i \hbar}e^{i \Omega_{\beta
\alpha}t''}v_{\beta \alpha}(t'')+\cdots
\label{2.7}
\eeq
The amplitude for a transition from $\Phi_\alpha$ at $t=-\infty$ to
$\Phi_\gamma$ at $t=\infty$ is $a_\gamma(\infty)$. Thus the perturbation 
series expansion of the transition amplitude can be written
\bea
a_\gamma(\infty)&=&\delta_{\gamma,\alpha}+\sum_{n=1}^\infty
\sum_{\beta\ldots\tau}\int_{-\infty}^\infty \frac{d t_n}{i
\hbar}e^{i \Omega_{\gamma \beta}t_n}v_{\gamma\beta}(t_n)\int_{-\infty}^{t_n
} \frac{d t_{n-1}}{i
\hbar}e^{i \Omega_{\beta \lambda}t_{n-1}}v_{\beta\lambda}(t_{n-1})\ldots
\label{2.8}\\
&&\ldots\int_{-\infty}^{t_{p+1}}\frac{d t_p}{i
\hbar}e^{i \Omega_{\mu \nu}t_p}v_{\mu\nu}(t_p)\ldots\int_{-\infty}^{t_3} 
\frac{d t_2}{i\hbar}e^{i \Omega_{\sigma \tau}t_2}v_{\sigma\tau}(t_2)
\int_{-\infty}^{t_2} \frac{d t_1}{i\hbar}e^{i 
\Omega_{\tau \alpha}t_1}v_{\tau\alpha}
(t_1)\nonumber
\eea
\section{Specialization to the quantum theory of relativistic 
Coulomb excitation}
In the usual approach to relativistic Coulomb excitation (RCE), the target
nucleus is at rest in the reference frame. The rapidly moving projectile 
nucleus follows a prescribed orbit, which we will leave unspecified. 
The projectile interacts with the electromagnetic field via its time-dependent 
charge and current densties 
$\rho_{{\rm P}}({\bf r},t)$ and ${\bf J}_{{\rm P}}({\bf r},t)$.

The Hamiltonian $h_0$ of Equation (\ref{2.1}) will refer to the target 
internal degrees of freedom, plus the free electromagnetic field. 
An eigenstate of $h_0$ will therefore be specified by a target state 
$\phi_\beta$, plus a specification of the numbers
of photons in each of the quantized modes of the field. If the energy of 
the target state $\phi_\beta$ is $\epsilon_\beta$, then a transition from 
target state $\phi_\alpha$ to $\phi_\beta$ with creation of a photon of 
momentum $\hbar{\bf q}$ will result in an energy increase of
$$
(\epsilon_\beta + \hbar c q)-\epsilon_\alpha,
$$
so that the quantity $\Omega_{\beta \alpha}$ referred to in Equation 
(\ref{2.8}) is
$$
\Omega_{\beta \alpha}=\frac{(\epsilon_\beta + \hbar c q)-\epsilon_\alpha}
{\hbar}=\frac{\epsilon_\beta-\epsilon_\alpha}{\hbar}+cq~\equiv~\omega_
{\beta \alpha}+cq.
$$
Similarly, if the $\phi_\alpha\rightarrow \phi_\beta$ transition were 
accompanied by the absorption of a photon of momentum $\hbar{\bf q}$, 
we would have
$$
\Omega_{\beta \alpha}=\frac{(\epsilon_\beta - \hbar c q)-\epsilon_\alpha}
{\hbar}=\frac{\epsilon_\beta-\epsilon_\alpha}{\hbar}-cq~\equiv~\omega_
{\beta \alpha}-cq.
$$

Since the calculation involves the electromagnetic potentials, we must
choose a gauge. In this section and the next, we use the \textit{Coulomb},
or \textit{radiation}, gauge, \cite{HEI,SAK,JAC} because it allows the 
simplest treatment of the quantized electromagnetic field. In Section V we 
will discuss the modifications 
required for the Lorentz gauge. In the Coulomb gauge, the vector potential is
a solenoidal field
\beq
{\bf \nabla \cdot A(r,t)}=0.
\label{3.1}
\eeq
Here ${\bf r}$ labels points of space relative to an origin at the
target center, which we assume to be fixed. We describe the photon field
in terms of normal modes, each labelled by a wave vector ${\bf q}$. 
Associated with each ${\bf q}$ are \textit{two} polarization vectors
$\eps_{{\bf q},1}$ and $\eps_{{\bf q},2}$. The three vectors 
$({\bf {\hat q}},~\eps_{{\bf q},1},~\eps_{{\bf q},2})$ form an orthonormal
coordinate system. The field operator ${\bf A(r})$ is expressed in
terms of photon creation and annihilation operators $\left(a^+_
{{\bf q},j}, a_{{\bf q},j}\right)$ by the expansion
\beq
{\bf A(r})=c\sqrt{\frac{4 \pi}{V}}~\sum_{\bf
q}\sum_{j=1}^2~\sqrt{\frac{\hbar}{2 q c}}~\left[~e^{i{\bf q \cdot r}}\eps
_{{\bf q},j} a_{{\bf q},j} ~+~e^{-i{\bf q \cdot r}}\eps_{{\bf q},j} ^* 
a^+_{{\bf q},j}~\right].
\label{3.2}
\eeq
Here $V$ is the quantization volume. The number of modes per unit 
${\bf q}$-space volume is $V/8 \pi^3$, for each polarization 
$\eps_{{\bf q},j}$. 
Since the plane waves that describe the
modes are transverse $(\eps_{{\bf q},j}
\cdot{\hat {\bf q}}=0)$, the solenoidal condition (\ref{3.1}) is 
automatically satisfied.
With the normalization given in equation (\ref{3.2}), the
$\left(a^+_{{\bf q},j}, a_{{\bf q},j}\right)$ obey the commutation
relations
\bes
\beq
[~a_{{\bf q},j}, a^+_{{\bf q'},j'}~]=\delta_{{\bf q},
{\bf q'}}\delta_{j,j'}
\label{3.3a}
\eeq
\beq
[~a_{{\bf q},j}, a_{{\bf q'},j'}~]=[~a^+_{{\bf q},j}, 
a^+_{{\bf q'},j'}~]=0
\label{3.3b}
\eeq
\ees

Their matrix elements with respect to states with $n_{{\bf q},j}
$ photons in the mode ${\bf q},j$ are
\bes
\bea
<n'_{{\bf q},j}~|~a^+_{{\bf q},j}~|~n_{{\bf q},j}>&=&\delta_{n'_{
{\bf q},j},n_{{\bf q},j}+1}\sqrt{n_{{\bf q},j}+1}\label{3.4a} \\
<n'_{{\bf q},j}~|~ a_{{\bf q},j}~|~n_{{\bf q},j}>&=&\delta_{n'_{
{\bf q},j},n_{{\bf q},j}-1}\sqrt{n_{{\bf q},j}}
\label{3.4b}
\eea
\ees

In the Coulomb gauge, the full Hamiltonian for our system is 
\cite{SAK,HEI}
\bes
\bea
H&=&h_0-\frac{1}{c}\int d^3r\left(~{\bf J}_{\rm P}
({\bf r},t)+{\bf J}_{\rm T}({\bf r})~\right)\cdot{\bf A(r)}+\int d^3r
\rho_{{\rm T}}({\bf r})\int d^3 r'\frac{\rho_{{\rm P}}
({\bf r'},t)}{|{\bf r - r'}|}\label{3.5a} \\
h_0&=&H_{{\rm T}}+H_\gamma
\label{3.5b}
\eea
\ees
$H_{{\rm T}}$ is the target internal Hamiltonian, and
$$
H_\gamma \equiv \sum_{\bf q}\sum_{j=1}^2~\hbar c q a^+_{{\bf q},j} 
a_{{\bf q},j}
$$
is the free-field photon Hamiltonian. There is no term in equation
(\ref{3.5a}) corresponding to the kinetic energy of relative motion. We
are using the usual RCE picture of the projectile moving on a prescribed
classical trajectory, and therefore the projectile-target
relative coordinate is not one of the degrees of freedom of the problem.
The projectile charge and current densities
($\rho_{{\rm P}}({\bf r'},t),~{\bf J}_{\rm P}({\bf r'},t)$) are to be 
regarded as specified functions of ${\bf r'},t$, whereas the target 
charge and current densities are to be regarded as operator functions, to be 
represented in the calculation by their matrix elements 
$[\rho_{{\rm T}} ]_{\gamma \beta}({\bf r},t) ,~[{\bf J}_{\rm } ]_
{\gamma,\beta}({\bf r},t) $ with 
respect to the target eigenstates $\phi_\gamma,\phi_\beta$. 

The interaction to be used in Equation (\ref{2.1}) is
\beq
v(t)=-\frac{1}{c}\int d^3r\left[~\left(~{\bf J}_{\rm P}
({\bf r},t)+{\bf J}_{\rm T}({\bf r})~\right)\cdot{\bf A(r)}~\right]+\int d^3r
\rho_{{\rm T}}({\bf r})\int d^3 r'\frac{\rho_{{\rm P}}
({\bf r'},t)}{|{\bf r - r'|}}
\label{3.6}
\eeq
It is conveniently regarded as a sum of two terms:
\bes
\beq
v_0(t)\equiv\int d^3r \rho_{{\rm T}}({\bf r})\int d^3 r'\frac{\rho_{{\rm P}}
({\bf r'},t)}{|{\bf r - r'}|} 
\label{3.7a}
\eeq
which does not change the number of photons, and
\beq
v_1(t)\equiv-\frac{1}{c}\int d^3r\left[~\left(~{\bf J}_{\rm P}
({\bf r},t)+{\bf J}_{\rm T}({\bf r})~\right)\cdot{\bf A(r)}~\right] 
\label{3.7b}
\eeq
\ees
which changes the number of photons by $\pm 1$ because of the presence of 
$a_{{\bf q},j}$ and
$a^+_{{\bf q'},j'}$ in the expansion of ${\bf A(r},t)$ (Equation (\ref{3.2})).

We are interested in a situation in which neither the initial state
$\phi_\alpha$ nor the final state $\phi_\gamma$ has any photons\footnote{To
 keep the notation simple, we use $\phi_{\alpha,\gamma}$ to label target 
states, or target states in the presence of the photon vacuum. We believe 
that the presence or absence of the vacuum will be clear in all situations.}. 
Thus we exclude bremsstrahlung processes. Moreover, we assume that
every photon absorbed by the target at time $t$ was emitted by the
projectile at an earlier time $t'$, and every photon emitted by the
target at time $t$ will be absorbed by the projectile at a later time
$t'$ \footnote{In particular, we do not imagine that a photon absorbed
by the target was emitted by the target. Such target emission-absorption
processes contribute to the renormaliztion of the target mass, and are
not considered here.}. This implies that the creation and annihilation 
operators entering into the expansion (Equation (\ref{2.8})) for the transition
amplitude will occur in pairs $a_{{\bf q},j}\ldots a^+_{{\bf q},j}.$ 
Specifically, if the $t_p$ integrand
 in Equation (\ref{2.8}) contained
$$
\frac{e^{i \Omega_{\mu \nu}t_p}}{i \hbar}~\eps_{{\bf q},j} \cdot 
[{\bf J}_{{\rm T}}]_{\mu \nu}({\bf r}) ~ a_{{\bf q},j}e^{i {\bf q \cdot r}}
~=~ \frac{e^{i (\omega_{\mu \nu}-cq)t_p}}{i \hbar}~\eps
_{{\bf q},j}\cdot [{\bf J}_{{\rm T}}]_{\mu \nu}({\bf r}) ~ a_{{\bf q},j}
e^{i {\bf q \cdot r}}
$$
corresponding to absorption of a $({\bf q},j)$ photon at the target at time
 $t_p$, it must be that this photon was created at the projectile at an 
earlier time $t'_p$. This means that farther to the right in the multiple 
integral of Equation (\ref{2.8}) there occurs a factor
$$
\frac{e^{i c q t'_p}}{i \hbar}~\eps_{{\bf q},j}^* \cdot {\bf J}_{{\rm P}}
({\bf r'},t'_p)~ a^+_{{\bf q},j}e^{-i {\bf q \cdot r'}}
$$
and $t'_p$ must be integrated from $-\infty$ to $t_p$. Similarly, if the 
$t_p$ integrand in Equation (\ref{2.8}) contained
$$
\frac{e^{i \Omega_{\mu \nu}t_p}}{i \hbar}~\eps _{{\bf q},j}^*\cdot [{\bf J}
_{{\rm T}}]_{\mu \nu}({\bf r}) ~ a^+_{{\bf q},j}e^{-i {\bf q \cdot r}}~=~
 \frac{e^{i (\omega_{\mu \nu}+cq)t_p}}{i \hbar}~\eps
_{{\bf q},j}^*\cdot [{\bf J}_{{\rm T}}]_{\mu \nu}({\bf r}) ~a^+_{{\bf q},j}
e^{-i {\bf q \cdot r}}
$$
corresponding to creation of a $({\bf q},j)$ photon at the target at time 
$t_p$, it must be that this photon will be annihilated at the projectile at 
a later time $t'_p$. This means that farther to the left in the multiple 
integral of Equation (\ref{2.8})
 there occurs a factor
$$
\frac{e^{-i c q t'_p}}{i \hbar}~\eps_{{\bf q},j} \cdot {\bf J}_{{\rm P}}
({\bf r'},t'_p)~a_{{\bf q},j}e^{+i {\bf q \cdot r'}}
$$
and $t'_p$ must be integrated from $t_p$ to $\infty$. The result of these 
two $t'_p$ integrations is that the double integral over $t_p$ and $t'_p$ in
 Equation (\ref{2.8}) is replaced by the single integral
\beq
\int_{-\infty}^{t_{p+1}}\frac{d t_p}{i \hbar}e^{i \omega_{\mu \nu} t_p}~
w_{\mu \nu}(t_p),
\label{3.8}
\eeq
where $w_{\mu \nu}(t_p)$ is defined by
\bea
w_{\mu \nu}(t_p)& \equiv &\frac{4 \pi}{V}\sum_{\bf q}\sum_{j=1}^2\frac
{\hbar}{2 q c} \label{3.9}\\
&~&\Bigl[~\int d^3r
\eps_{{\bf q},j}^*\cdot [{\bf J}_{{\rm T}}]_{\mu \nu}({\bf r})
e^{i(cqt_p-{\bf q  \cdot r})}\int_{t_p}^\infty\frac{d t'_p}{i \hbar}\int d^3r'
\eps_{{\bf q},j}\cdot {\bf J}_{{\rm P}}({\bf r'},t'_p)
e^{i(-cqt'_p+{\bf q \cdot r'})}\nonumber\\
&+& \int d^3r
\eps_{{\bf q},j}\cdot [{\bf J}_{{\rm T}}]_{\mu \nu}({\bf r})
e^{i(-cqt_p+{\bf q \cdot r})}\int_{-\infty}^{t_p}\frac{d t'_p}{i \hbar}
\int d^3r'
\eps_{{\bf q},j}^*\cdot {\bf J}_{{\rm P}}({\bf r'},t'_p)
e^{i(cqt'_p-{\bf q \cdot r'})}~\Bigr]\nonumber
\eea
Note that it is possible to group all the contributions to the $t'_p$ range
 in Equation (\ref{2.8}) into these two intervals, $-\infty \leq t'_p \leq 
t_p$ and $\infty \geq t'_p \geq t_p$, only because of our assumption that 
the projectile current density
is independent of what occurs at the projectile or target.

We can repeat this process for every $a_{{\bf q},j}\ldots a^+_{{\bf q},j}$ 
pair ocurring in the multiple integral of Equation (\ref{2.8}), thereby 
replacing every $t,t'$ integration ($t$ ocurring at the target, $t'$ at the 
projectile) by a single $t$ integration. All that will remain in 
Equation (\ref{2.8}) is integrals of the form (\ref{3.8}) obtained in this way,
and integrals of the form
$$
\int_{-\infty}^{t_{\ell+1}}\frac{d t_\ell}{i \hbar}e^{i \omega_{\mu \nu}
t_\ell}v_0(t_\ell).
$$
Note that here $\Omega_{\mu \nu}=\omega_{\mu \nu}$, since $v_0(t_\ell)$ 
does not create or annihilate photons. The remaining combination of the 
integrals
$$
\int_{-\infty}^{t_{p+1}}\frac{d t_p}{i \hbar}e^{i \omega_{\mu \nu} t_p}
w_{\mu \nu}(t_p)~~~{\rm and}~~~\int_{-\infty}^{t_{\ell+1}}\frac{d t_\ell}
{i \hbar}e^{i \omega_{\mu \nu}t_\ell}v_0(t_\ell)
$$
is equal to
\bea
a_\gamma(\infty)&=&\delta_{\gamma,\alpha}+\sum_{n=1}^\infty\sum_{\beta
\ldots\tau}\int_{-\infty}^\infty \frac{d t_n}
{i\hbar}e^{i \omega_{\gamma \beta}t_n}{\tilde v} _{\gamma\beta}(t_n)
\int_{-\infty}^{t_n} \frac{d t_{n-1}}
{i\hbar}e^{i \omega_{\beta \lambda}t_{n-1}}{\tilde v} _{\beta\lambda}
(t_{n-1})\ldots\label{3.10}\\
&&\ldots\int_{-\infty}^{t_{p+1}}\frac{d t_p}
{i\hbar}e^{i \omega_{\mu \nu}t_p}{\tilde v} _{\mu\nu}(t_p)\ldots
\int_{-\infty}^{t_3} \frac{d t_2}{i\hbar}e^{i \omega_{\sigma \tau}t_2}
{\tilde v} _{\sigma\tau}(t_2)\int_{-\infty}^{t_2} \frac{d t_1}{i\hbar}
e^{i \omega_{\tau \alpha}t_1}{\tilde v} _{\tau\alpha}(t_1)\nonumber
\eea
in which the time-dependent perturbation ${\tilde v}(t)$ is defined by
\beq
{\tilde v} (t)~\equiv~v_0(t)~+~w(t).
\label{3.11}
\eeq
Comparison with Equation (\ref{2.8}) shows that the transition amplitude 
for the excitation of target state $\phi_\gamma$ under the influence of the 
quantized electromagnetic field is precisely the same as if the target had 
experienced the time dependent effective field ${\tilde v}(t)$, from which 
all photon degrees of freedom have been removed.

In the next Section, we will discuss the physical meaning of the effective 
target interaction ${\tilde v}(t)$ defined in Equation (\ref{3.11}).

The structure of Equation (\ref{3.10}) can be elucidated by examining the 
$n=2$ term in detail:
\beq
a_\gamma^{(2)}(\infty)=\sum_\beta\int_{-\infty}^\infty\frac{dt_2}
{i \hbar}e^{i \omega_{\gamma \beta}t_2}\left[v_0(t_2)+w(t_2)\right]_
{\gamma \beta}\int_{- \infty}^{t_2}\frac{d t_1}{i \hbar}
e^{i \omega_{\beta \alpha}t_1}\left[v_0(t_1)+w(t_1)\right]_{\beta \alpha}
\label{3.12}
\eeq
The $v_0(t_2)v_0(t_1)$ combination arises from the $n=2$ term of Equation
 (\ref{2.8}), when the $v_0$ part of $v(t)$ (Equation (\ref{3.7a})) is used 
both times. The $v_0(t_2)w(t_1)$ and $w(t_2)v_0(t_1)$ combinations arise 
from the $n=3$ term of Equation ({\ref{2.8}), when one of the three 
interactions is chosen to be $v_0$ and the other two are chosen to be 
photon interactions $v_1$ (Equation (\ref{3.7b})). The $w(t_2)w(t_1)$ 
combination arises
 from the $n=4$ term of Equation (\ref{2.8}), with $v_1$ acting four times.
This involves the exchange of two photons between projectile and target,
 the target interactions occurring at $t_2$, $t_1$ and the corresponding 
projectile interactions occurring at $t_2',t_1'$, with both $t_2'$ and $t_1'$
integrated from $-\infty$ to $\infty$. These components of the $n=2$ 
term of Equation (\ref{3.10}) are illustrated in Figure 1,2 and 3.
\section{The effective interaction }
If we define four fields ${\tilde \varphi}({\bf r},t)$ and 
${\tilde {\bf A}}({\bf r},t)$ by
\bes
\bea
{\tilde \varphi}({\bf r},t)&\equiv&\int d^3r'\frac{\rho_{{\rm P}}({\bf r'},
t)}{|{\bf r - r'}|} \label{4.1a} \\
{\tilde {\bf A}}({\bf r},t)&\equiv&-\frac{2 \pi}{iV}\sum_{\bf q}
\sum_{j=1}^2\frac{1}{q}
 \Bigl[~
\eps_{{\bf q},j}e^{i(cqt-{\bf q  \cdot r})}\int_{t}^\infty\frac{d t'}
{i \hbar}\int d^3r'
\eps_{{\bf q},j}\cdot [{\bf J}_{{\rm P}}]({\bf r'},t')
e^{i(-cqt'+{\bf q \cdot r'})}\nonumber\\
&+&\eps_{{\bf q},j}e^{i(-cqt+{\bf q \cdot r})}\int_{-\infty}^{t}\frac{d t'}
{i \hbar}\int d^3r'
\eps_{{\bf q},j}^*\cdot [{\bf J}_{{\rm P}}]({\bf r'},t')
e^{i(cqt'-{\bf q \cdot r'})}~\Bigr],\label{4.1b}
\eea
\ees
then we can re-write Equations (\ref{3.11}, \ref{3.9}, and \ref{3.7a}) as 
follows:
\beq
{\tilde v}(t)=\int d^3r~\left[~\rho_{{\rm T}}({\bf r}){\tilde \varphi}
({\bf r},t)~-~\frac{1}{c}{\bf J}_{{\rm T}}({\bf r},t)\cdot{\tilde {\bf A}}
({\bf r},t)~\right].
\label{4.2}
\eeq
It follows immediately from these definitions of ${\tilde \varphi}({\bf r},
t)$ and ${\tilde {\bf A}}({\bf r},t)$ that
\bes
\bea
\nabla^2{\tilde \varphi}({\bf r},t)&=&-4\pi\rho_{{\rm P}}({\bf r},t) 
\label{4.3a}\\
\nabla \cdot {\tilde {\bf A}}({\bf r},t)&=&0 \label{4.3b}.
\eea
\ees
The latter equation is a consequence of our use of transverse photon modes 
$(\eps_{{\bf q},j}\cdot {\bf {\hat q}}=0= \eps_{{\bf q},j}^*\cdot 
{\bf {\hat q}})$.

To proceed, we replace the sum $\sum_{{\bf q}}$ by the integral 
$\frac{V}{8\pi^3}\int d^3q$.
Moreover the completeness of the set $(\eps_{{\bf q},1}, \eps_{{\bf q},2},
{\bf {\hat q}})$ allows us to write
$$
{\bf J}_{{\rm P}}={\bf {\hat q}}~({\bf {\hat q}}\cdot{\bf J}_{{\rm P}})
~+~ \sum_{j=1}^2\eps_{{\bf q},j}^*~(\eps_{{\bf q},j}\cdot 
{\bf J}_{{\rm P}})
$$
from which we can obtain
$$
\sum_{j=1}^2\eps_{{\bf q},j}^*~(\eps_{{\bf q},j}\cdot {\bf J}_{{\rm P}})
=\sum_{j=1}^2\eps_{{\bf q},j}~(\eps_{{\bf q},j}^*\cdot 
{\bf J}_{{\rm P}})={\bf J}_{{\rm P}}-{\bf {\hat q}}~({\bf {\hat q}}
\cdot {\bf J}_{{\rm P}})={\bf J}_{{\rm P}} -\frac{{\bf q}({\bf q}\cdot 
{\bf J}_{{\rm P}})}{q^2}.
$$
Then we can rewrite Equation (\ref{4.1b}) as
\bea
{\tilde {\bf A}}({\bf r},t)&=&\frac{i}{4 \pi^2}\int\frac{d^3q}{q}
 \Bigl[~e^{i(cqt-{\bf q  \cdot r})}\int_{t}^\infty d t'\int d^3r'
\left({\bf J}_{{\rm P}} ({\bf r'},t') -\frac{{\bf q}({\bf q}\cdot 
{\bf J}_{{\rm P}}({\bf r'},t') )}{q^2}~\right) e^{i(-cqt'+{\bf q \cdot r'})}
\nonumber\\
&+&
e^{i(-cqt+{\bf q \cdot r})}\int_{-\infty}^{t}d t'\int d^3r'
\left({\bf J}_{{\rm P}} ({\bf r'},t') -\frac{{\bf q}({\bf q}\cdot 
{\bf J}_{{\rm P}}({\bf r'},t') )}{q^2}~\right)
e^{i(cqt'-{\bf q \cdot r'})}\Bigr].\label{4.4}
\eea
By carrying out the differentiations, one can show that
\bes
\beq
\nabla^2{\tilde {\bf A}}({\bf r},t)-\frac{1}{c^2}
\frac{\partial^2{\tilde {\bf A}}({\bf r},t)}{\partial t^2}=
\frac{-1}{2 \pi^2c}\int d^3q\int d^3r'\left({\bf J}_{{\rm P}} ({\bf r'},t) 
-\frac{{\bf q}({\bf q}\cdot {\bf J}_{{\rm P}}({\bf r'},t) )}{q2}~\right) 
e^{i {\bf q \cdot (r-r')}}
\label{4.5a}
\eeq
and
\beq\frac{\partial}{\partial t}\nabla {\tilde \varphi}({\bf r},t)=\frac{1}
{2 \pi^2}\int \frac{d^3q}{q^2}i{\bf q} \int d^3r'e^{i {\bf q \cdot (r-r')}}
\frac{\partial \rho_{{\rm P}}({\bf r'},t)}{\partial t}.
\label{4.5b}
\eeq
But projectile charge conservation implies that
$$
\int d^3r' e^{i {\bf q \cdot (r-r')}} \frac{\partial \rho_{{\rm P}}
({\bf r'},t)}{\partial t}=-\int d^3r'e^{i {\bf q \cdot (r-r')}} 
\nabla_{{\bf r'}} \cdot{\bf J}_{{\rm P}}({\bf r'},t)
$$
$$
=\int d^3r' {\bf J}_{{\rm P}}({\bf r'},t)\cdot\nabla_{{\bf r'}}
e^{i {\bf q \cdot (r-r')}}=-i\int d^3r'{\bf q \cdot J}_{{\rm P}}
({\bf r'},t)e^{i {\bf q \cdot (r-r')}}.
$$
Thus Equation (\ref{4.5b}) becomes
\beq
\frac{\partial}{\partial t}\nabla {\tilde \varphi}({\bf r},t)=
\frac{1}{2 \pi^2}\int d^3q \int d^3r' \frac{{\bf q}
({\bf q \cdot J}_{{\rm P}}({\bf r'},t))}{q^2} e^{i {\bf q \cdot (r-r')}},
\label{4.5c}
\eeq
\ees
which we can combine with Equation (\ref{4.5a}) to get
$$
\nabla^2{\tilde {\bf A}}({\bf r},t)-\frac{1}{c^2}
\frac{\partial^2{\tilde {\bf A}}({\bf r},t)}{\partial t^2}-\frac{1}{c}
\frac{\partial}{\partial t}\nabla {\tilde \varphi}({\bf r},t)=-\frac{1}
{2 \pi^2c}\int d^3r'{\bf J}_{{\rm P}}({\bf r'},t)\int d^3q 
e^{i {\bf q \cdot (r-r')}})
$$
$$
=-\frac{1}{2 \pi^2c}\int d^3r'~(2\pi)^3\delta({\bf r-r'})
{\bf J}_{{\rm P}}({\bf r'},t)
$$
\beq
\nabla^2{\tilde {\bf A}}({\bf r},t)-\frac{1}{c^2}\frac{\partial^2{\tilde 
{\bf A}}({\bf r},t)}{\partial t^2}-\frac{1}{c}
\frac{\partial}{\partial t}\nabla {\tilde \varphi}({\bf r},t)=
-\frac{4 \pi}{c}{\bf J}_{{\rm P}}({\bf r},t)
\label{4.6}
\eeq

Equations (\ref{4.3a}, \ref{4.3b} and \ref{4.6}) show that ${\tilde{\bf A}}
({\bf r},t)$ and ${\tilde \varphi}({\bf r},t)$ defined in Equations 
(\ref{4.1a}, \ref{4.1b}) are the classical vector and scalar potentials 
${\bf A}({\bf r},t)$ and $\varphi({\bf r},t)$ associated (in Coulomb gauge) 
with projectile charge and current densities 
$\left(\rho_{{\rm P}}({\bf r},t),{\bf J}_{{\rm P}}({\bf r},t)\right).$ 
Morevover, Equation (\ref{4.2}) shows that 
the effective potential ${\tilde v}(t)$ is the classical interaction of 
$\left(\varphi({\bf r},t),{\bf A(r},t)\right)$ with the target charge and 
current densities.

Thus we have proven that an RCE calculation in which the classical 
electromagnetic field of the projectile is used to induce transitions between 
target states gives the same target transition amplitudes, to all orders of 
perturbation theory, as would a calculation in which the interaction between 
projectile and target is mediated by a quantized electromagnetic field.

\section{The Lorentz gauge}

The two previous sections used the Coulomb gauge. In this section, we will
describe the modifications needed if the Lorentz gauge is used.
\begin{enumerate}
\item In the Lorentz gauge, the full Hamiltonian is
\beq
H=h_0+\int d^3r\left(~\left( \rho_{{\rm P}}({\bf r},t)+\rho_{{\rm T}}
({\bf r})\right)\varphi({\bf r}) -\frac{1}{c} \left(~{\bf J}_{\rm P}
({\bf r},t)+{\bf J}_{\rm T}({\bf r})~\right)\cdot{\bf A(r)}~\right),
\label{5.1}
\eeq
which lacks the density-density interaction present in Equation 
(\ref{3.5a}).
\item In the Coulomb gauge, the vector potential ${\bf A}$ has only two 
components, and they are transverse. In the Lorentz gauge, ${\bf A}$ has these
 two tranverse components, and also a longitudinal component (along 
${\bf {\hat q}}$). Moreover, the scalar potential ${\bf \varphi}$ is also 
quantized. Thus, in addition to the two transverse photons of the Coulomb 
gauge, we have a longitudinal photon and a scalar photon. The commutation 
relations of the longitudinal photon creation and annihilation operators are 
the same as for the transverse photons (Equations (\ref{3.3a}) and (\ref{3.3b})). 
However, the treatment of the scalar photon is more complicated. A consistent 
formalism for quantizing the scalar field was developed by 
S.N. Gupta\cite{Gu50} and  K. Bleuler\cite{Bl50}. For our purposes, the only 
manifestation of the extra complications of scalar field quantization is the 
presence of the two minus signs in Equation (V.2) below.
\end{enumerate}
When these changes are accounted for, Equation (\ref {3.9}) is replaced by
\bea
{\tilde v}_{\mu \nu}(t_p)&=&w_{\mu \nu}(t_p)=\frac{4 \pi}{V}
\sum_{\bf q}\frac{\hbar}{2 q c} \label{5.2}\\
&~&\Bigl[~\sum_{j=1}^3\int d^3r
\eps_{{\bf q},j}^*\cdot [{\bf J}_{{\rm T}}]_{\mu \nu}({\bf r})
e^{i(cqt_p-{\bf q  \cdot r})}\int_{t_p}^\infty\frac{d t'_p}{i \hbar}
\int d^3r'\eps_{{\bf q},j}\cdot {\bf J}_{{\rm P}}({\bf r'},t'_p)
e^{i(-cqt'_p+{\bf q \cdot r'})}\nonumber\\
&+& \sum_{j=1}^3\int d^3r
\eps_{{\bf q},j}\cdot [{\bf J}_{{\rm T}}]_{\mu \nu}({\bf r})e^{i(-cqt_p+
{\bf q \cdot r})}\int_{-\infty}^{t_p}\frac{d t'_p}{i \hbar}\int d^3r'
\eps_{{\bf q},j}^*\cdot {\bf J}_{{\rm P}}({\bf r'},t'_p)
e^{i(cqt'_p-{\bf q \cdot r'})} \nonumber\\
&-&c^2\int d^3r
[\rho_{{\rm T}}]_{\mu \nu}({\bf r})e^{i(cqt_p-{\bf q  \cdot r})}
\int_{t_p}^\infty\frac{d t'_p}{i \hbar}\int d^3r'
\rho_{{\rm P}}({\bf r'},t'_p)e^{i(-cqt'_p+{\bf q \cdot r'})}\nonumber\\
&-& c^2\int d^3r
[\rho_{{\rm T}}]_{\mu \nu}({\bf r})e^{i(-cqt_p+{\bf q \cdot r})}
\int_{-\infty}^{t_p}\frac{d t'_p}{i \hbar}\int d^3r'
\rho_{{\rm P}}({\bf r'},t'_p)e^{i(cqt'_p-{\bf q \cdot r'})}~\Bigr]\nonumber
\eea
Following the procedure of Section 4, we cast this expression in the form 
(\ref{4.2}), which leads to
\bes
\bea
{\tilde \varphi}({\bf r},t)&\equiv&-\frac{2c \pi}{iV}\sum_{\bf q}
\sum_{j=1}^3\frac{1}{q}
 \Bigl[~e^{i(cqt-{\bf q  \cdot r})}\int_{t}^\infty dt'\int d^3r'
\rho_{{\rm P}}({\bf r'},t')e^{i(-cqt'+{\bf q \cdot r'})}\nonumber\\
&+&e^{i(-cqt+{\bf q \cdot r})}\int_{-\infty}^{t}dt' \int d^3r'
\rho_{{\rm P}}({\bf r'},t')e^{i(cqt'-{\bf q \cdot r'})}~\Bigr]\label{5.3a}\\
{\tilde {\bf A}}({\bf r},t)&\equiv&-\frac{2 \pi}{iV}\sum_{\bf q}
\sum_{j=1}^3\frac{1}{q}
 \Bigl[~\eps_{{\bf q},j}^*e^{i(cqt-{\bf q  \cdot r})}\int_{t}^\infty dt' 
\int d^3r'
\eps_{{\bf q},j}\cdot [{\bf J}_{{\rm P}}]({\bf r'},t')
e^{i(-cqt'+{\bf q \cdot r'})}\nonumber\\
&+&
\eps_{{\bf q},j}e^{i(-cqt+{\bf q \cdot r})}\int_{-\infty}^{t} dt'\int d^3r'
\eps_{{\bf q},j}^*\cdot [{\bf J}_{{\rm P}}]({\bf r'},t')
e^{i(cqt'-{\bf q \cdot r'})}~\Bigr]\label{5.3b}
\eea
\ees
The polarization sum simplifies to
$$
\sum_{j=1}^3\eps_{{\bf q},j}^*~(\eps_{{\bf q},j}\cdot {\bf J}_{{\rm P}})
={\bf J}_{{\rm P}},
$$
and the replacement of $\sum_{{\bf q}}$ by the integral $\frac{V}{8\pi3}
\int d^3q$ still applies. When these relations are used, it is staightforward 
to calculate that ${\tilde \varphi}({\bf r},t)$ and ${\tilde {\bf A}}
({\bf r},t)$ defined by Equations (\ref{5.3a} and {\ref{5.3b}) satisfy
\bes
\bea
\left[~\nabla^2-\frac{1}{c^2}\frac{\partial^2}{\partial t^2}~\right]
{\tilde \varphi}({\bf r},t)&=&-4 \pi \rho_{{\rm P}}({\bf r},t)\label{5.4a}\\
\left[~\nabla^2-\frac{1}{c^2}\frac{\partial^2}{\partial t^2}~\right]{\tilde
 {\bf A}}({\bf r},t)&=&-\frac{4 \pi}{c} {\bf J}_{{\rm P}}({\bf r},t)
\label{5.4b}\\
\nabla \cdot {\tilde {\bf A}}({\bf r},t)+\frac{1}{c}\frac{\partial 
{\tilde \varphi}({\bf r},t)}{\partial t}&=&0 \label{5.4c}
\eea
\ees

Equations (\ref{5.4a}, \ref{5.4b} and \ref{5.4c}) show that 
${\tilde \varphi}({\bf r},t)$ and ${\tilde {\bf A}}({\bf r},t)$ defined 
in Equations (\ref{5.3a}, \ref{5.3b}) are the classical vector and 
scalar potentials ${\bf A}({\bf r},t)$ and $\varphi({\bf r},t)$ associated 
(in Lorentz gauge) with projectile charge and current densities 
$\left(\rho_{{\rm P}}({\bf r},t),{\bf J}_{{\rm P}}({\bf r},t)\right).$ Thus, 
whether one chooses to work in the 
Coulomb or Lorentz gauge, in an RCE calculation of target transition 
amplitudes the effect of the quantized electromagnetic field can be replaced 
by the classical electromagnetic field of the projectile used in the 
classical expression \ref{4.2} for the interaction.

The proof of the equivalence of the quantized field and classical field 
treatments of Coulomb excitation given in ABHMW \cite{ABHMW} applies only to 
the on-shell ($\omega=\omega_{\gamma \alpha})$ Fourier component of 
${\tilde v}_{\gamma \alpha}(t)$. The reason for this apparent restriction 
is that ABHMW used the Coulomb gauge for the calculation of the 
quantized-field version of ${\tilde v}_{\gamma \alpha}(t)$, but they used 
the Lorentz gauge for the calculation of the classical field version. It is 
shown in Reference \cite{Newgauge} that the versions of 
${\tilde v}_{\gamma \alpha}(t)$ calculated in the two gauges agree in their 
on-shell Fourier components, but are generally different when 
$\omega \neq \omega_{\gamma \alpha}$. In Sections
 III, IV, and V, we have shown that the quantized field and classical field
 treatments agree for {\it all} $t$, and so for {\it all} 
$\omega$, if both are calculated in the same gauge.

\section{Discussion}
It has been emphasized that an important ingredient in our derivation is 
the assumption that the  charge and current densities associated with the 
projectile are specified functions of position and time, which is equivalent 
to the assumption that the projectile does not change its internal state 
during the collision. This assumption is also made in RCE calculations 
involving classical electromagnetic fields. Even if both the target and 
projectile 
are allowed to undergo internal transitions, the fields that induce those 
transitions are always assumed to be generated by moving static spherically 
symmetric charge distributions, with no internal structure. To remove this 
assumption, we would have to calculate, at each instant, the classical 
retarded electromagnetic fields that have been generated by the projectile 
and target since the beginning of the collision, while they have moved 
relative to each other and undergone changes of their internal states. 
These would 
be the electromagnetic fields that, in turn, would be used to induce 
further internal transitions. It is clear that this would be a very difficult 
calculation, even though it involved only classical fields. It would be the 
classical analogue of the quantized field calculation we have described in 
Sections II and III, if we had allowed the projectile state to change as 
well as the target state.

Although it would be difficult to perform the complete calculation just 
described, in which the electromagnetic fields are generated by the actual 
dynamic transition charge and current densities, it may be required for a 
full understanding of ultra-high-energy RCE. Baltz, Rhoades-Brown and Weneser 
\cite{BRW} have estimated excitation probabilites for collisions between 
oppositely directed 100 GeV Au nuclei. They find excitation probabilities 
that are greater than 0.5 for grazing collisons, and greater than 0.1 out to 
impact parameters of about 50 fm. Thus it is important to investigate, with 
classical and/or quantized electromagntic fields, the effect on excitation 
probabilities of internal transtions occurring within both the target and 
projectile.

\newpage

\centerline{Figure Captions}
\medskip

Figure 1. The $v_0(t_2)v_0(t_1)$ term of Equation (\ref{3.12}). 
Time increases as we move upwards in the diagram. The vertical line on the 
left corresponds to the projectile, and the lines labelled 
$\alpha, \beta,\gamma$ represent states of the target. 

Figure 2. The $w(t_1)v_0(t_2)$ and $v_0(t_1)w(t_2)$ terms in Equation 
(\ref{3.12}). The curly lines represent exchanged photons. The 
$t_1'$ and $t_2'$ variables must be integrated from $-\infty$ to $\infty$.

Figure 3. The $w(t_1)w(t_2)$ term of Equation (\ref{3.12}). The $t_1'$ and 
$t_2'$ variables must be integrated, independently, from $-\infty$ to 
$\infty$.

\newpage
\includegraphics[scale=0.9]{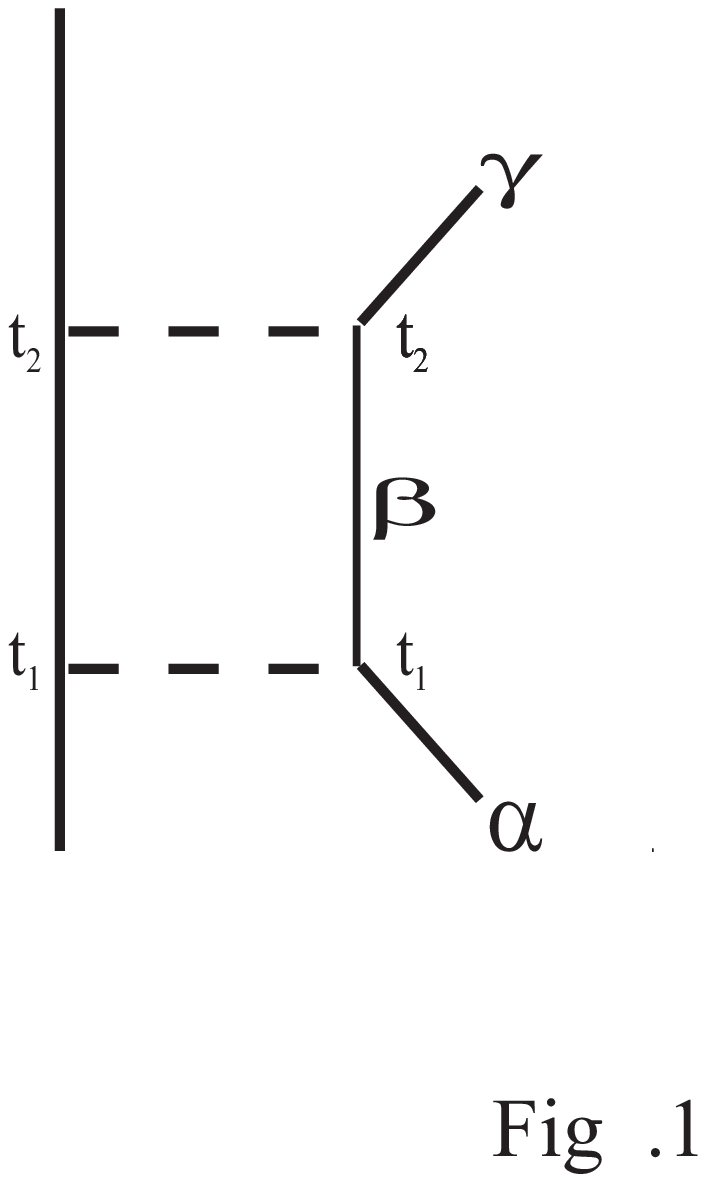}
\newpage
\includegraphics[scale=0.9]{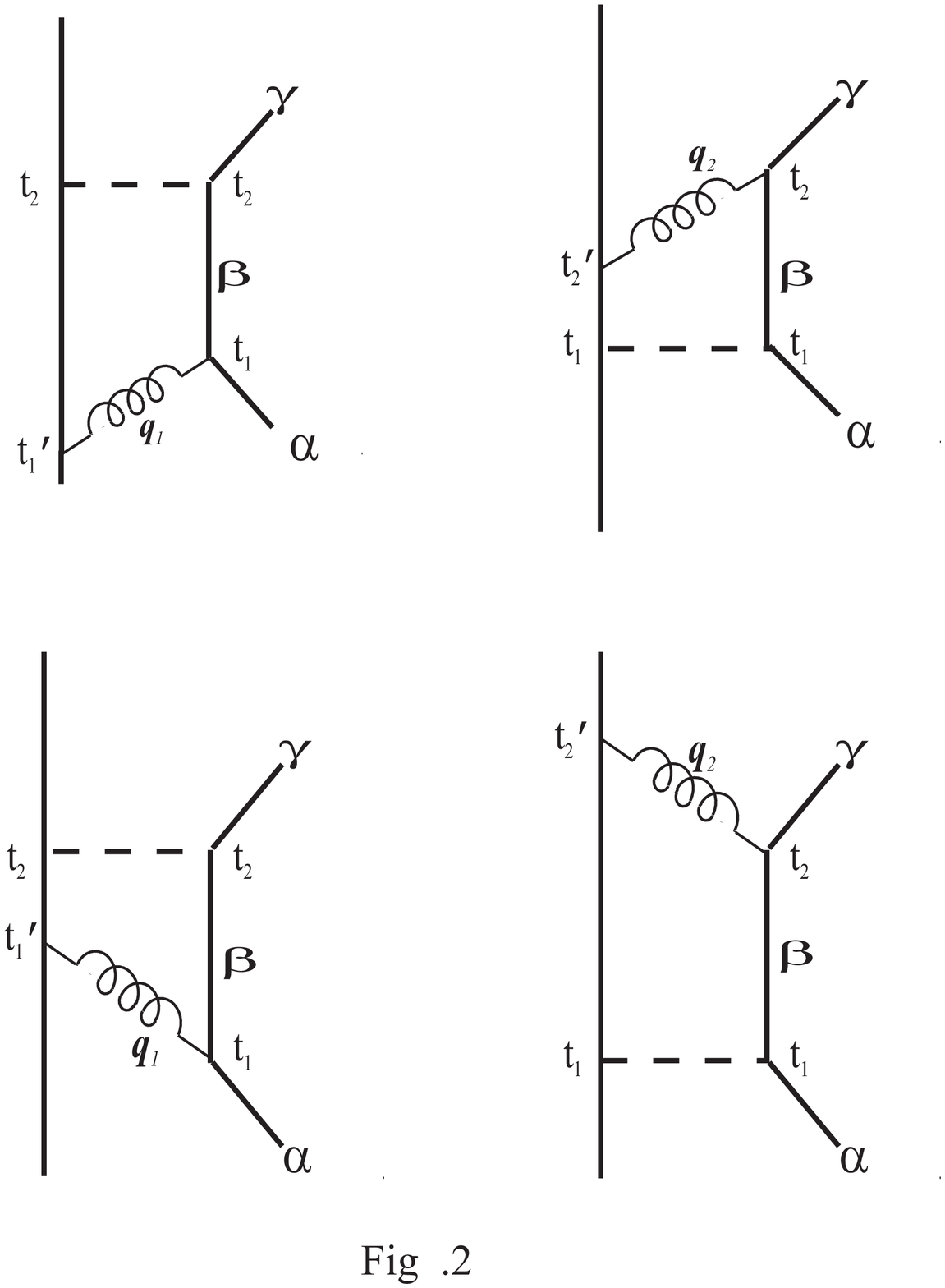}
\newpage
\includegraphics[scale=0.9]{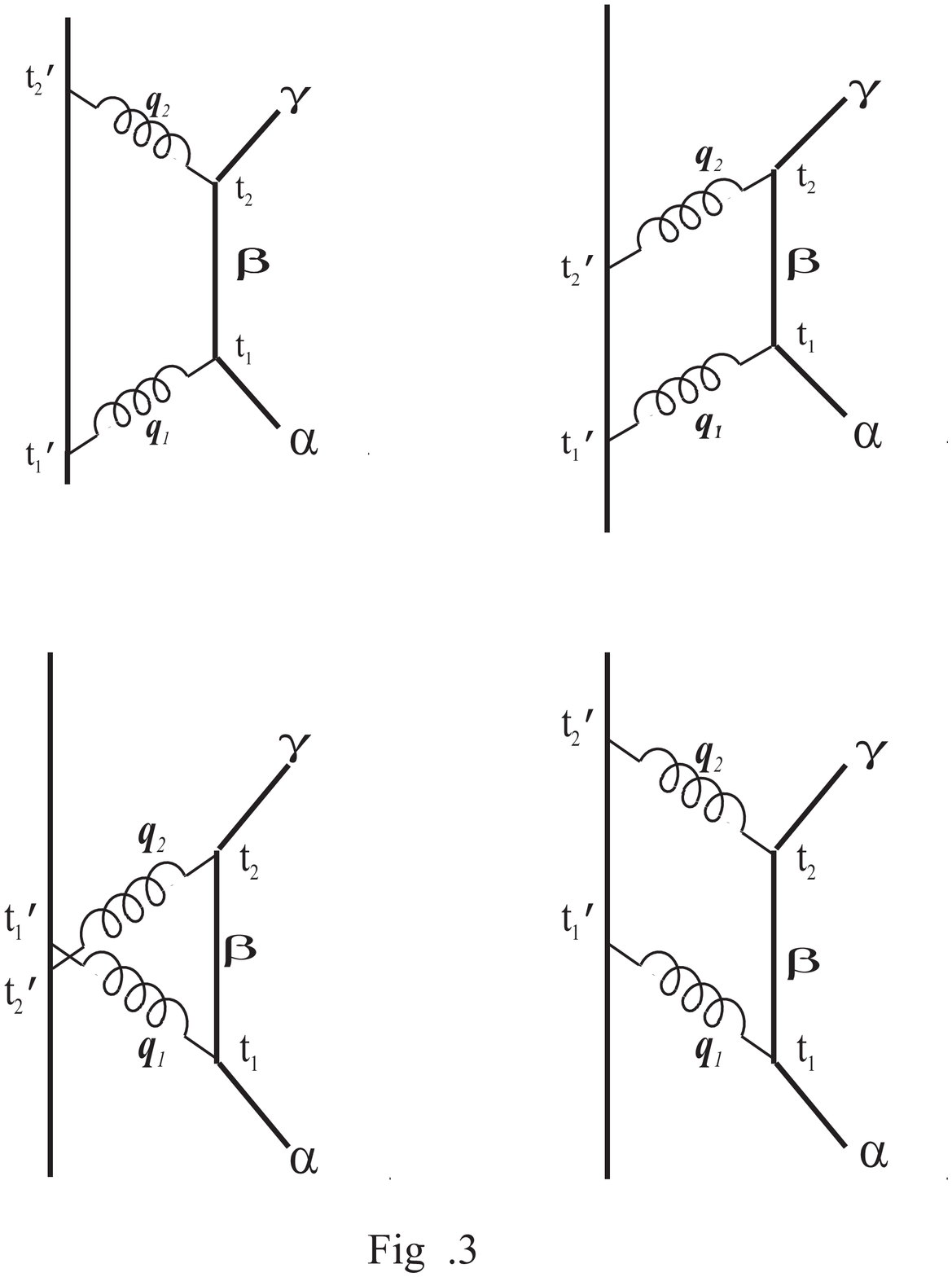}
\end{document}